\documentclass[fleqn,usenatbib,lineno]{mnras}

\usepackage{lineno}
\usepackage{newtxtext, newtxmath, nicefrac}
\usepackage[T1]{fontenc}
\usepackage{tikz}
\usepackage{hyperref}
\usepackage{xcolor}
\definecolor{orcidlogocol}{HTML}{A6CE39}

\DeclareRobustCommand{\VAN}[3]{#2}
\let\VANthebibliography\thebibliography
\def\thebibliography{\DeclareRobustCommand{\VAN}[3]{##3}\VANthebibliography}

\usepackage{graphicx}
\usepackage{amsmath}

\newcommand{\txd}{{\text{d}}}

\newcommand{\calE}{{\cal{E}}}

\newcommand{\calS}{{\cal{S}}}

\newcommand{\rT}{r_{\scriptscriptstyle{\text{T}}}}

\newcommand{\calET}{\calE_{\scriptscriptstyle{\text{T}}}}

\newcommand{\orcidicon}{%
    \tikz[baseline=-1.2ex]\node[fill=orcidlogocol, circle, inner sep=0.6pt]
        {\includegraphics[width=0.5ex]{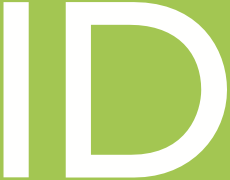}};}
\newcommand{\orcid}[1]{\href{https://orcid.org/#1}{\orcidicon}}

\DeclareMathOperator{\logit}{logit}
\DeclareMathOperator{\atanh}{atanh}
\DeclareMathOperator{\erf}{erf}

\defcitealias{Baes2022b}{Paper~I}
\defcitealias{Baes2023a}{Paper~II}
\defcitealias{Baes2023b}{Paper~III}
\defcitealias{Baes2024c}{Paper~IV}

\title[Dynamical models with a finite extent~V]{Self-consistent dynamical models with a finite extent -- V. Smooth radial truncations and phase-space consistency}

\author[M. Baes]{Maarten Baes\,\orcid{0000-0002-3930-2757}\,{}%
\thanks{E-mail: maarten.baes@ugent.be}
\\Sterrenkundig Observatorium, Universiteit Gent, Krijgslaan 299, 9000 Gent, Belgium
}

\date{\today}

\pubyear{2025}

\begin{document}
\label{firstpage}
\pagerange{\pageref{firstpage}--\pageref{lastpage}}
\maketitle

\begin{abstract}
Many stellar systems exhibit a finite spatial extent, yet constructing self-consistent spherical models with a prescribed outer boundary is non-trivial because sharp density cutoffs introduce discontinuities that lead to inconsistencies in the associated distribution function. In this paper we show that these difficulties arise from the abruptness of the truncation rather than from the finite extent itself. We introduce a general and infinitely differentiable radial truncation scheme that can be applied to any density profile, and illustrate its behaviour using the Hernquist model. We find that softly truncated models are dynamically consistent provided that the truncation is sufficiently gradual, and we determine the corresponding critical truncation sharpness. Their distribution functions display a characteristic bump--dip feature near the truncation energy that signals the transition between consistent and inconsistent cases. In contrast to sharply truncated models, softly truncated systems can support an extensive family of Osipkov--Merritt orbital structures, including moderately radial ones. Soft truncations therefore offer a general and physically motivated route to constructing finite-extent dynamical models with well-controlled outer-edge behaviour.
\end{abstract}


\section{Introduction}

Analytical dynamical models remain essential in galaxy dynamics, serving as testbeds for novel methods and as foundations for numerical simulations. Simple spherical models such as the Plummer sphere \citep{Plummer1911, Dejonghe1987}, the Jaffe model \citep{Jaffe1983}, the Hernquist model \citep{Hernquist1990}, the $\gamma$-models \citep{Dehnen1993, Tremaine1994}, and the Einasto models \citep{Einasto1965} are widely used for a large variety of applications. One common characteristic of these models is that they have infinite spatial extent, whereas real systems such as galaxies and star clusters have a finite radial extent. This immediately raises the question whether it is possible to convert these popular models into counterparts with a finite extent or, more generally, to construct self-consistent dynamical models with a finite extent.

One method to generate simple dynamical models with a finite extent is to apply a truncation in binding energy, which automatically excludes low-binding-energy orbits beyond a truncation radius. However, this approach introduces two issues. First, it artificially excludes certain orbits, even though they are completely within the allowed spatial region defined by the truncation radius \citep{Kashlinsky1988}. Secondly, a truncation in binding energy affects all dynamical properties. Even when the density and potential of the original model are simple analytical functions, the corresponding quantities of the truncated model can virtually never be written down in closed form. The prototypical example of such models is the family of King models \citep{King1966}. The distribution function has a simple expression when written as a function of binding energy, but no closed form exists for the density or gravitational potential.

This paper is the fifth in a series that aims at studying the dynamical structure of radially truncated models and at developing self-consistent spherical dynamical models with finite extent and preset analytical density profiles. In \citetalias{Baes2022b} \citep{Baes2022b}, we started our exploration with the simplest finite-extent model: the uniform density sphere. We confirmed that this model cannot be supported by an isotropic dynamical structure (see also \citealt{Zeldovich1972, Osipkov1979}), but we developed a range of self-consistent dynamical models in which all permissible orbits are fully populated. In \citetalias{Baes2023a} \citep{Baes2023a}, we demonstrated that spherical models with a sharp density cutoff cannot sustain an isotropic orbital structure. Many of them can, however, be supported by the tangential Osipkov--Merritt orbital structure that becomes entirely tangential at the truncation radius. In \citetalias{Baes2023b} \citep{Baes2023b}, we examined the detailed properties of the family of power-law spheres, for which many dynamical properties can be calculated analytically. In \citetalias{Baes2024c} \citep{Baes2024c} we presented a new family of finite-extent models defined by density profiles based on Wendland functions, the most widely used compactly supported radial basis functions \citep{Wendland1995, Schaback2011}. We demonstrated that these models, which have an analytical potential--density pair, can be supported by a range of orbital structures. While Wendland models are not representative of real astrophysical systems, the main insight from this study was that models with truncated density profiles can be supported by isotropic and even radially anisotropic orbital structures as long as the truncation is sufficiently smooth.

In this paper we combine the lessons learnt from the previous papers in this series and aim at constructing dynamical models with a finite extent by applying a soft radial truncation to an arbitrary density profile. In Sect.~{\ref{construction.sec}} we present our methodology and the numerical implementation in the {\tt{SpheCow}} code. In Sect.~{\ref{Hernquist.sec}} we apply this method to the well-known Hernquist model and investigate the properties of a family of softly truncated Hernquist models. In particular, we demonstrate that this model is consistent with a range of orbital structures, ranging from tangentially anisotropic to isotropic to radially anisotropic. In Sect.~{\ref{discussion.sec}} we discuss the implications of our results, and in Sect.~{\ref{summary.sec}} we provide a summary.

\section{Construction of softly truncated models}
\label{construction.sec}

\subsection{Mathematical formulation}
\label{Math.sec}

In \citetalias{Baes2023a} we discussed models with an infinitely sharp radial truncation. More specifically, we started from an arbitrary spherical density profile $\rho_0(r)$ with infinite extent, thus $\rho_0(r) > 0$ for all $r$. We then created a new model by applying a radial truncation to this density profile,
\begin{equation}
\rho(r) = \rho_0(r)\, \Theta(\rT - r),
\label{sharptruncation}
\end{equation}
with $\rT$ the truncation radius and $\Theta(x)$ the Heaviside step function,
\begin{equation}
\Theta(x) = \begin{cases} 
\;0 & \quad {\text{for }} x < 0, \\ 
\;1 & \quad {\text{for }} x > 0. 
\end{cases} 
\end{equation}
As shown in \citetalias{Baes2023a}, such models cannot be supported by an ergodic distribution function. However, we also argued in that paper that replacing the sharp truncation with a soft one might be the key to generate dynamically consistent radially truncated models. We therefore replace the sharp truncation~(\ref{sharptruncation}) by a soft one,
\begin{equation}
\rho(r) = \rho_0(r)\, \calS(y),
\qquad
y = \frac{1 - r/\rT}{1 - \xi},
\label{rho}
\end{equation}
where the truncation function $\calS(x)$ is a monotonically increasing and sufficiently smooth function that satisfies
\begin{equation}
\calS(x) = \begin{cases} 
\;0 & \quad {\text{for }} x \leqslant 0, \\ 
\;1 & \quad {\text{for }} x \geqslant 1. 
\end{cases} 
\label{conditionS}
\end{equation}
The truncation sharpness $\xi$ is a parameter with $0\leqslant\xi<1$ (the limiting case $\xi = 1$ corresponds to the limiting sharp truncation, recovered when the width of the truncation region tends to zero). The density profile $\rho(r)$ is equivalent to the original density profile $\rho_0(r)$ for $r \leqslant \xi\,\rT$. For $\xi\,\rT < r < \rT$, it gradually deviates from the original density profile until it smoothly converges to zero at the truncation radius $\rT$. We hence have a soft truncation that sets in at $r=\xi\,\rT$ and that terminates at $r=\rT$. As the name indicates, the truncation sharpness parameter controls the width or sharpness of the truncation region: small $\xi$ yields a gradual, extended softening, whereas large $\xi$ produces a sharper transition.

There is no unique functional form for the truncation function $\calS(x)$. To find a suitable candidate, we consider a continuous probability density function $\phi(x)$ supported on the closed interval $[0,1]$. If we then consider the cumulative density function (CDF) $\Phi(x)$ and set
\begin{equation}
\calS(x) = \begin{cases} 
\;0 & \quad {\text{for }} x \leqslant 0, \\ 
\;\Phi(x) & \quad {\text{for }} 0 < x < 1, \\
 \;1 & \quad {\text{for }} x \geqslant 1, 
\end{cases} 
\label{defS}
\end{equation}
we obtain a monotonically increasing function that satisfies the condition~(\ref{conditionS}). To ensure that the truncation function is sufficiently smooth, the probability density function should be chosen such that its lowest-order derivatives at $x=0$ and $x=1$ vanish. 

A convenient probability distribution is the logit--normal distribution, widely used in statistics and neural networks \citep{Aitchison1980, Lenk1988}. For the standard parameters $\mu=0$ and $\sigma=1$, it is characterised by the probability density
\begin{equation}
\phi(x) = \frac{1}{\sqrt{2\pi}}\,\frac{1}{x\,(1-x)}\,{\text{e}}^{-\frac12 (\logit x)^2},
\end{equation}
with $\logit$ the inverse of the standard logistic function,
\begin{equation}
\logit x =  \ln\left( \frac{x}{1-x} \right) = 2 \atanh (2x-1).
\end{equation}
The logit--normal distribution has CDF
\begin{equation}
\Phi(x) = \frac12\left[ 1 + \erf \left(\frac{\logit x}{\sqrt2}\right) \right] .
\label{logit-normal-CDF}
\end{equation}
A characteristic that is important for our purposes is that all derivatives of the logit--normal CDF vanish at $x=0$ and $x=1$, or in other words, the function~(\ref{defS}) corresponding to the logit--normal CDF~(\ref{logit-normal-CDF}) is infinitely differentiable over the entire real line.

\begin{figure}
\includegraphics[width=0.95\columnwidth]{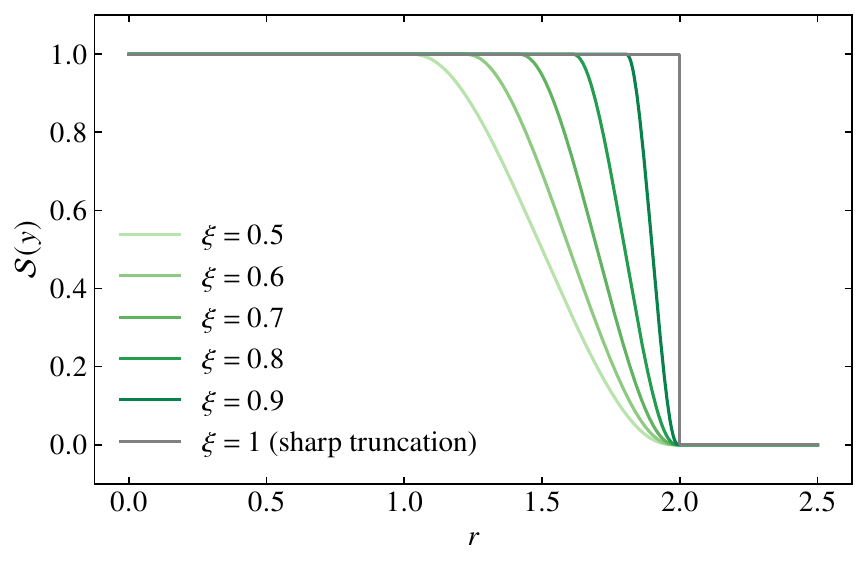}%
\caption{Illustration of the truncation function $\calS(y)$. The truncation radius is fixed at $\rT = 2$, whereas the truncation sharpness takes different values ranging from 0.5 to 1.}
\label{truncationfunction.fig}
\end{figure}

Fig.~{\ref{truncationfunction.fig}} illustrates the behaviour of the logit--normal truncation function for $\rT=2$ and different values of $\xi$. This figure shows the role of the truncation sharpness parameter: small $\xi$ yields a gradual, extended softening, while large $\xi$ produces a sharper transition. In the limiting case $\xi = 1$ the truncation becomes infinitely sharp. For alternative choices of the truncation function we refer to Appendix~{\ref{appendix.sec}}.

We note that the truncation method described here, as the case of the sharp truncations discussed in \citetalias{Baes2023a}, is not mass-conservative: the total mass of the truncated models is always reduced compared to the original model. A mass-conservative scheme can be set up by truncating the original density profile and redistributing the truncated mass over the remaining part. In this case, we add a multiplicative factor to Eq.~(\ref{rho}), 
\begin{equation}
\rho(r) = W\,\rho_0(r)\,{\cal{S}}(y)
\end{equation}
with
\begin{equation}
W = \dfrac{\int_0^\infty \rho_0(r)\,r^2\,\txd r}{\int_0^\infty \rho_0(r)\,{\cal{S}}(y)\,r^2\,\txd r}.
\end{equation}
In the remainder of this work, we choose the non-mass-conservative scheme (\ref{rho}). Since the difference between both schemes comes down to a weight factor in the density profile and all other relevant dynamical properties, the main results obtained are insensitive of the choice of the scheme.

\begin{figure*}
\centering
\includegraphics[width=\textwidth]{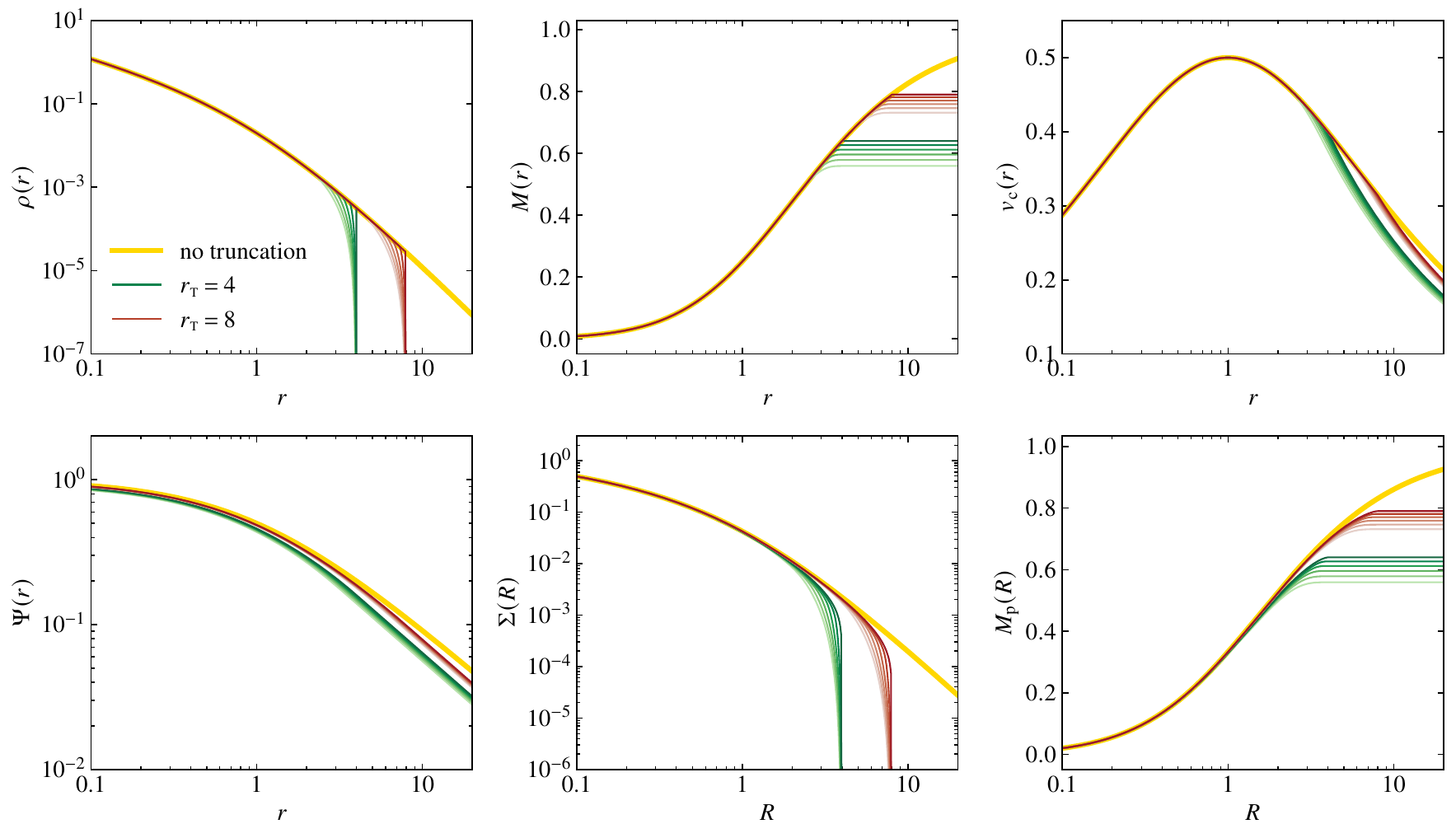}%
\caption{Basic properties of softly truncated Hernquist models. Properties shown are the density, mass profile, circular velocity curve, gravitational potential, surface density, and projected mass profile. The thick golden line corresponds to the standard Hernquist model without truncation; the thinner lines correspond to softly truncated models with different values for the truncation radius $\rT$ and the truncation sharpness $\xi$. The latter ranges from 0.5 (lightest shade) to 1 (darkest shade) in steps of 0.1. We adopt normalised units with $G = M = b = 1$.}
\label{Hernquist_basic.fig}
\end{figure*}

\subsection{Implementation in {\tt{SpheCow}}}

{\tt{SpheCow}} \citep{Baes2021b} is a versatile tool designed to explore the dynamical properties of spherical models. The code comes with implementations for many popular mass models, defined by either their density or their surface density profiles. Its architecture is user-friendly, allowing new models to be added with minimal effort. Users can assign an ergodic, purely radial, or radial or tangential Osipkov--Merritt orbital structure to any mass model and numerically explore the entire dynamical structure by means of dynamical properties such as the velocity dispersion profiles, the distribution function, and the differential energy distribution. {\tt{SpheCow}} is written in C++ and is publicly available on GitHub.\footnote{\url{https://github.com/mbaes/SpheCow}}

We implemented the option to generate softly truncated versions, with user-specified values for $\rT$ and $\xi$, for any {\tt{SpheCow}} model. Similarly to the sharply truncated models discussed in \citetalias{Baes2023a}, we employed the decorator design pattern, a well-known object-oriented programming pattern that allows additional functionality to be dynamically added to objects \citep{Gamma1994, Freeman2004}. This approach avoids the need to build truncated versions for every existing and future density model and ensures that soft truncations can be applied to any model, including ones without analytical density profiles, such as the S\'ersic and Nuker models.

The implementation of the softly truncated model decorator class in {\tt{SpheCow}} is straightforward and efficient. Indeed, for any model defined by a density profile, the code only requires an implementation of the density and its first and second derivatives. Given the density profile $\rho_0(r)$ of the base model that is to be truncated and the values of the parameters $\rT$ and $\xi$ that set the truncation, the density $\rho(r)$ is obtained by combining expressions~(\ref{rho}), (\ref{defS}) and (\ref{logit-normal-CDF}). The first and second derivatives are immediately obtained and easily implemented, given the expressions for $\rho_0'(r)$ and $\rho_0''(r)$, which are available for the base model. All other intrinsic properties, projected properties, and dynamical properties are automatically calculated using high-order Gauss--Legendre quadrature \citep[for details, see][]{Baes2021b}. The additional evaluation of the smoothing function and its derivatives is negligible compared to the repeated quadrature dominating the evaluation of the dynamical quantities.

\section{Dynamical structure of softly truncated Hernquist models}
\label{Hernquist.sec}

\subsection{The Hernquist model}

To investigate the effect of a soft truncation on the structure of dynamical models, we use the Hernquist model \citep{Hernquist1990} as our starting point. This model is characterised by the simple density profile
\begin{gather}
\rho_0(r) = \frac{M}{2\pi\,b^3} \left(\frac{r}{b}\right)^{-1} \left(1+\frac{r}{b}\right)^{-3},
\end{gather}
with $M$ the total mass and $b$ a scale length. As nearly all popular models, it has an infinite extent, thus $\rho_0(r) > 0$ for all $r$. The Hernquist model is one of the most popular toy models in the stellar dynamics community: it can be supported by a variety of orbital structures, and the corresponding distribution functions and differential energy distributions can often be expressed completely analytically \citep{Hernquist1990, Ciotti1996, Baes2002, Baes2021c, Binney2008}.

\begin{figure*}
\centering
\includegraphics[width=\textwidth]{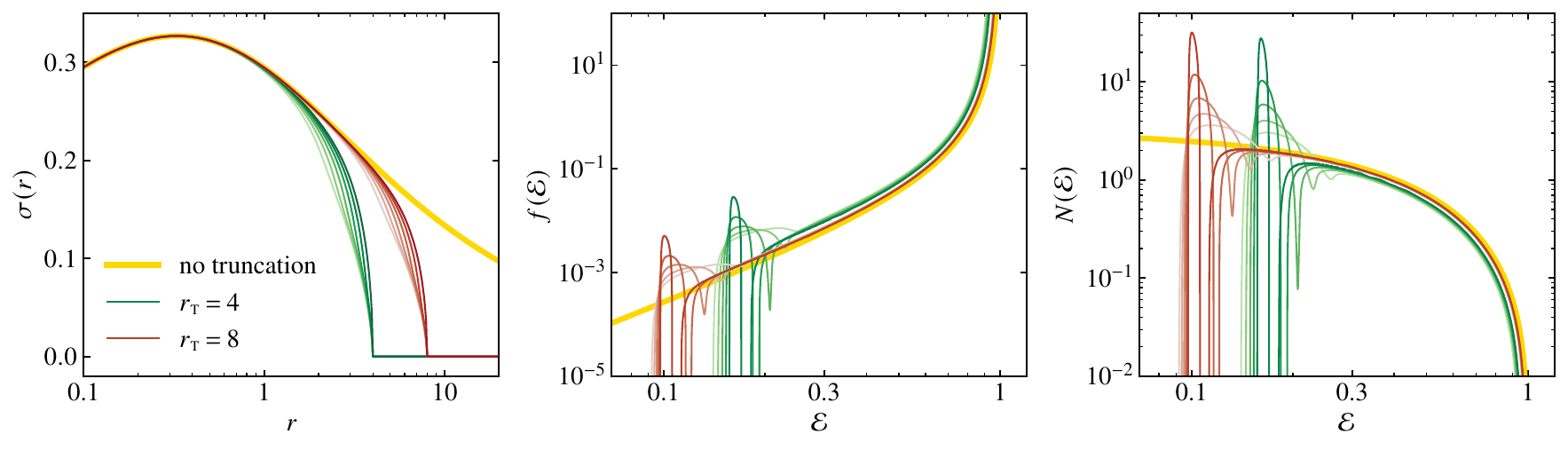}%
\caption{Dynamical properties of softly truncated Hernquist models with an isotropic orbital structure. Properties shown are the velocity dispersion profile, phase-space distribution function, and differential energy distribution. The different lines have the same meaning as in Fig.~{\ref{Hernquist_basic.fig}}. For the sharp truncation model ($\xi=1$), the distribution function and differential energy distribution are not shown.}
\label{Hernquist_iso.fig}
\end{figure*}

\subsection{Basic properties}

Fig.~{\ref{Hernquist_basic.fig}} shows a number of basic properties of softly truncated Hernquist models, for two different values of the truncation radius ($\rT = 4$ and 8) and different values of the truncation sharpness $\xi$, ranging from 0.5 to 1. The effect of both parameters on the density profile (top-left panel) is as expected: instead of a density profile that decreases smoothly and remains positive for all $r$, we obtain a density profile that drops towards zero at the truncation radius. The higher the truncation sharpness parameter, the larger the radius at which the truncation sets in and the less smooth the transition. Owing to this truncation, the cumulative mass profile (top-middle panel) of the softly truncated models is identical to that of the original model up to the onset of the truncation at $r=\xi\,\rT$. From that radius onwards, the mass profile quickly converges to a value that is necessarily below the total mass of the original model. Similarly, the circular velocity curve (top-right panel) of the truncated models is not affected before the onset of the truncation and transforms to a purely Keplerian decline for $r\gtrsim\rT$. All quantities in the top row are, at fixed radius, monotonically increasing functions of $\rT$ and $\xi$.

The panels on the bottom row of Fig.~{\ref{Hernquist_basic.fig}} show, from left to right, the relative potential, the surface density profile, and the cumulative projected mass profile. As for the quantities shown in the top row, at fixed (projected) radius, all of these quantities are increasing functions of $\rT$ and $\xi$. A difference is that the truncation affects the quantities shown in the bottom row at every (projected) radius, even at (projected) radii smaller than $\xi\,\rT$. Indeed, each of these quantities depends on the density at all radii, including those at large radii. For example, the potential is not only affected at large radii, with a Keplerian decline beyond the truncation radius, but the depth of the potential well, $\Psi_0$, is also reduced by the truncation.

\subsection{Dynamical properties}
\label{DynamicalProperties.sec}

One of the main questions we want to address in this study is whether softly truncated models can be supported by an ergodic orbital structure. In Fig.~{\ref{Hernquist_iso.fig}} we show the most important dynamical properties of the same models as in Fig.~{\ref{Hernquist_basic.fig}}.

The velocity dispersion profile (left panel) can be calculated from the density and cumulative mass profile by solving the Jeans equation. At small radii, the dispersion profile is similar to, but not exactly equal to, the dispersion profile of the original Hernquist model. At larger radii, it starts to differ and subsequently drops towards zero at the truncation radius. We note that this difference already becomes substantial at radii smaller than the onset of the truncation at $r=\xi\,\rT$. 

A dynamical model is only physically meaningful if it is consistent, i.e. if the corresponding distribution function is nonnegative over the entire phase space. This consistency can be investigated by explicitly calculating the distribution function, which, in the case of ergodic dynamical models, depends only on the binding energy $\calE$. For any spherical potential--density pair, the unique ergodic distribution function $f(\calE)$ can be calculated through the Eddington formula \citep{Eddington1916, Binney2008}. The isotropic distribution function of the softly truncated Hernquist models is shown in the middle panel of Fig.~{\ref{Hernquist_iso.fig}}. 

While the distribution function of the Hernquist model asymptotically behaves as $f(\calE) \propto \calE^5$ for $\calE\to0$ \citep{Hernquist1990}, one expects a different behaviour for the truncated models. Indeed, since the density vanishes beyond $r=\rT$, all orbits with $\calE < \calET \equiv \Psi(\rT)$ have to be unpopulated, and thus $f(\calE) = 0$ for $\calE < \calET$. In all cases, the distribution function of the softly truncated models has a characteristic behaviour. At high binding energies it mimics the original Hernquist model, with a power-law behaviour $f(\calE) \propto (\Psi_0-\calE)^{-5/2}$. At binding energies corresponding approximately to the onset of the truncation, the distribution function shows a sudden decrease. This decrease does not continue, however: it is soon followed by an increase, such that a local dip is formed. The distribution function then shows a local maximum, a bump, and subsequently decreases towards zero at $\calE =  \calET$. For larger values of $\rT$, the bump--dip combination appears at smaller binding energies. For a fixed truncation radius, the dip and the peak of the bump shift towards smaller binding energies and become more pronounced as the truncation sharpness parameter increases.

An analogous behaviour is seen for the differential energy distribution (right panel), which corresponds to the mass per unit binding energy. For the original isotropic Hernquist model, the differential energy distribution converges to a finite value as $\calE\to0$ \citep{Hernquist1990}, while the truncated models all have a different energy distribution characterised by a binding-energy truncation at $\calE = \calET$ and a bump--dip combination at $\calE \gtrsim \calET$, which shifts towards lower binding energies and becomes more pronounced as the sharpness of the truncation increases. 

\begin{figure}
\hspace*{1em}
\includegraphics[width=0.83\columnwidth]{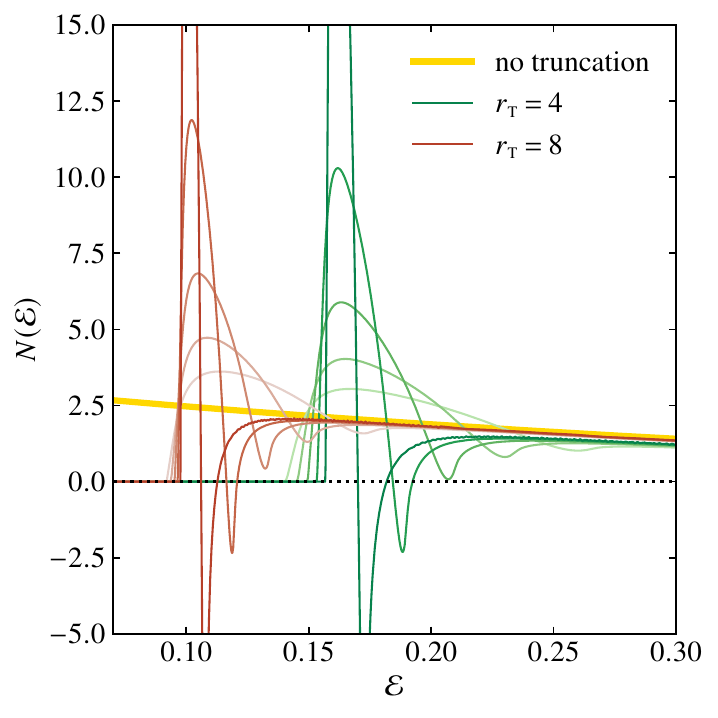}%
\caption{Zoom into the bump--dip region of the differential energy distribution of softly truncated Hernquist models with an isotropic orbital structure. The $y$-axis is shown in linear scaling. The different lines have the same meaning as in Figs.~{\ref{Hernquist_basic.fig}} and~{\ref{Hernquist_iso.fig}}.}
\label{Hernquist_iso_N.fig}
\end{figure}

\begin{table}
\caption{Critical truncation sharpness for the softly truncated Hernquist models. All models with $\xi < \xi_{\text{crit}}(\rT)$ can be supported by an isotropic orbital structure.}
\label{xicrit.tab}
\centering
\begin{tabular}{cc}
$\rT$ & $\xi_{\text{crit}}$ \\ \hline 
2 & 0.6723 \\
3 & 0.6940 \\
4 & 0.7061 \\
5 & 0.7137 \\
6 & 0.7189 \\
7 & 0.7228 \\
8 & 0.7257 \\
9 & 0.7280 \\
10 & 0.7299 \\
\hline
\end{tabular}
\end{table}

The key question is now whether the distribution function, or equivalently the differential energy distribution, is nonnegative for all binding energies. This depends on the values of the parameters $\rT$ and $\xi$. For relatively soft truncations, the value of the dip is modest and $f(\calE)$ and $N(\calE)$ are positive over the entire range $\calET < \calE < \Psi_0$, which means that the isotropic model is consistent. However, as the truncation sharpness increases, the dip grows deeper, and at some critical value $\xi_{\text{crit}}(\rT)$ it reaches zero. For truncation sharpness parameters $\xi > \xi_{\text{crit}}(\rT)$, $f(\calE)$ and $N(\calE)$ become negative at the minimum of the dip. This is clearly illustrated in Fig.~{\ref{Hernquist_iso_N.fig}}, which shows a zoom of $N(\calE)$ in the interesting range in binding energy, plotted with a linear $y$-axis. This figure shows that, for both $\rT = 4$ and $\rT=8$, the models with $\xi\leqslant0.7$ are consistent, whereas the models with $\xi\geqslant0.8$ are not. An in-depth investigation shows that $\xi_{\text{crit}}$ is a monotonically increasing function of $\rT$ (see Table~{\ref{xicrit.tab}}): models with larger truncation radii allow slightly sharper truncations.

\section{Discussion}
\label{discussion.sec}

\subsection{Sharp versus soft truncations}

The main result of this study is that spherically symmetric models with a finite spatial extent can be supported by an isotropic orbital structure, provided that the truncation of the density profile is sufficiently smooth. This finding stands in contrast to the main result of \citetalias{Baes2023a}, where we demonstrated that ergodic models with a sharp density cutoff are always inconsistent. The present results thus demonstrate that the failure of those models is not due to the finite extent itself, but to the discontinuous nature of the truncation.

The physical origin of this behaviour lies in the continuity of the phase-space occupation near the truncation energy. In a sharply truncated model, the density remains finite up to the truncation radius and drops to zero abruptly. Such a discontinuity in configuration space corresponds to a singularity in phase space: to reproduce the sudden drop, the distribution function must include delta-function or negative-weight contributions at the truncation energy. This results in a non-physical or negative distribution function, violating the requirement of dynamical self-consistency.

When the density profile is truncated in a smooth way so that it tends continuously to zero, this singularity disappears. The potential and the binding-energy distribution vary smoothly, and the occupancy of orbit families changes gradually rather than abruptly. All orbits can then be populated with finite, positive weights, and both the density and the velocity dispersion naturally vanish at the system boundary. The resulting models are dynamically consistent without the need for artificial anisotropies.

However, not every soft truncation guarantees consistency. If the truncation is only weakly smoothed, so that the density gradient near the outer radius remains extremely steep, the system effectively retains a quasi-discontinuity in its outer phase-space structure. In such cases, the derivatives of the density with respect to the potential still diverge as $r\to\rT$, and the corresponding distribution function can again become negative near the truncation energy. A sharp but finite truncation thus represents a marginally pathological case: less severe than an abrupt cutoff but still inconsistent because the density's high-order derivatives enforce an unphysical behaviour in energy space.\footnote{This intermediate regime mirrors the situation encountered in the transition from broken to double power-law models \citep{Zhao1996, Baes2021a}: only when the transition between the inner and outer slopes is sufficiently gradual does the ergodic distribution function remain positive. The same argument explains why Einasto models with Einasto indices $n<\tfrac12$ cannot support an isotropic orbital structure. Indeed, for these models the transition between the inner and outer regimes is too harsh \citep{Baes2022d}.} 

The successful construction of ergodic dynamical models by applying a soft truncation to the Hernquist density profile confirms that finite-extent systems can be realised without invoking discontinuous or extreme orbital configurations. The proposed soft-truncation prescription thus offers a general and physically motivated framework for generating self-consistent dynamical models with finite spatial extent and controlled outer-edge behaviour. It bridges the gap between idealised infinite models such as the popular Plummer, Dehnen, Einasto and S\'ersic models, and realistic stellar systems of finite size.

\begin{figure}
\includegraphics[width=0.95\columnwidth]{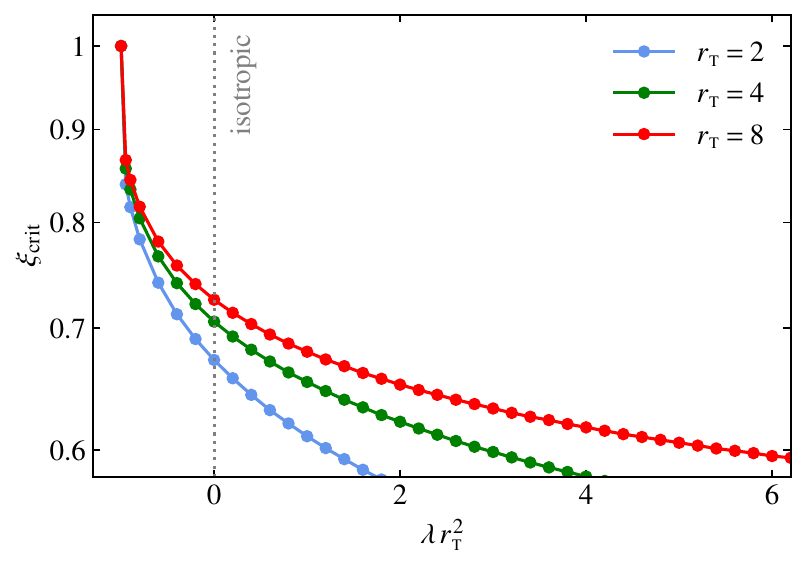}%
\caption{The critical truncation sharpness $\xi_{\text{crit}}(\rT,\lambda)$ for truncated Hernquist models with different truncation radii and different Osipkov--Merritt parameters. All Osipkov--Merritt models with $\xi\leqslant\xi_{\text{crit}}$ are consistent. The dotted grey line corresponds to $\lambda=0$, i.e. isotropic models.}
\label{xicrit.fig}
\end{figure}

\subsection{Extension to anisotropic orbital structures}
\label{OM.sec}

\begin{figure*}
\centering
\includegraphics[width=\textwidth]{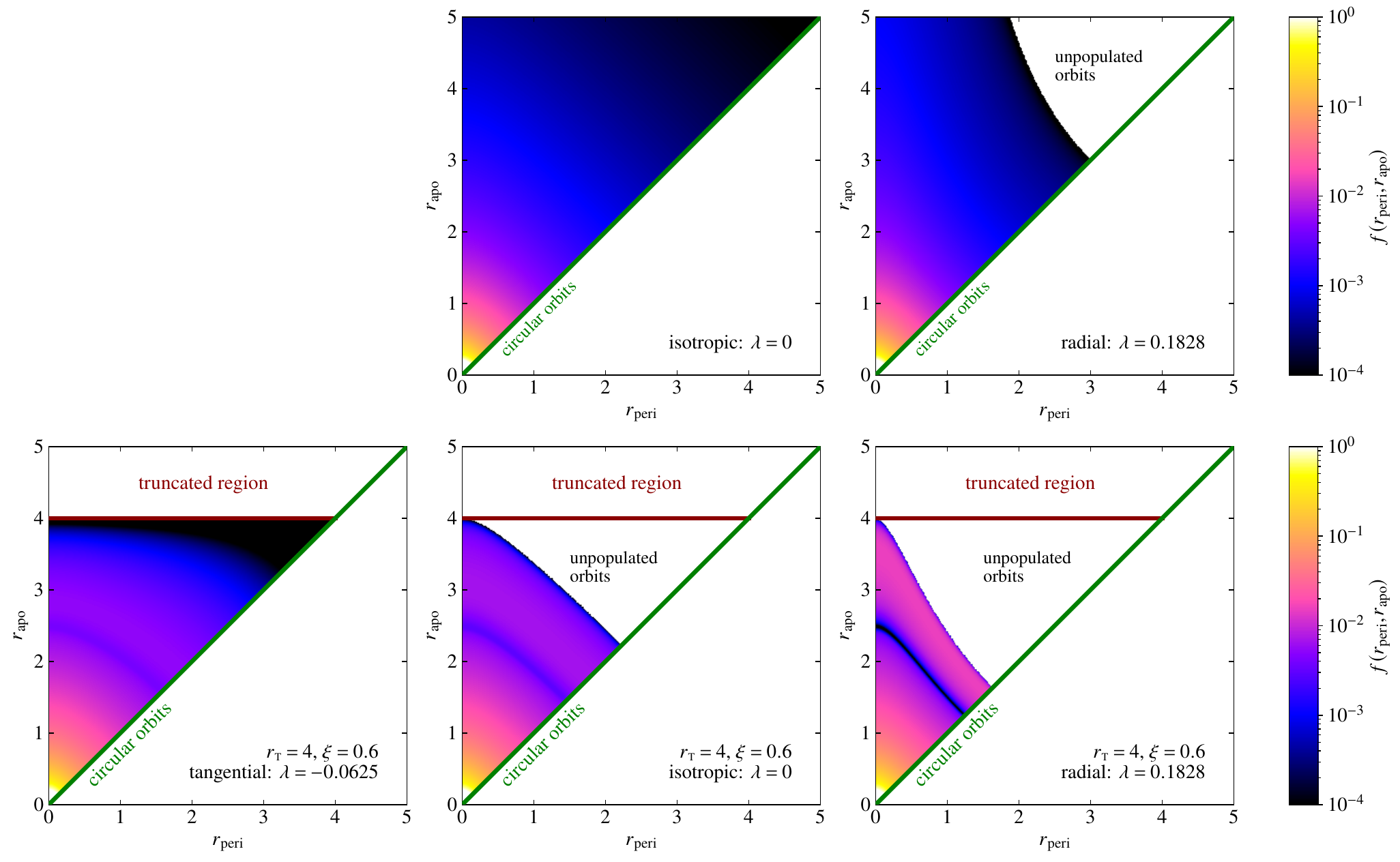}%
\caption{Distribution functions in turning-point space, $f(r_{\text{peri}}, r_{\text{apo}})$, for non-truncated and truncated Hernquist models. The top row corresponds to the non-truncated Hernquist model and shows the distribution function for the isotropic model and for the radial Osipkov--Merritt model with $\lambda=0.1828$. The bottom row corresponds to the truncated Hernquist model with $\rT=4$ and $\xi=0.6$. It shows the distribution functions, from left to right, for the most tangential Osipkov--Merritt model ($\lambda = \lambda_{\text{min}} = -0.0625$), the isotropic model, and the most radial Osipkov--Merritt model ($\lambda = \lambda_{\text{max}} = 0.1828$).}
\label{DFTP.fig}
\end{figure*}

So far we have focused on isotropic dynamical models, the simplest and most popular orbital structure. Observed dynamical systems have more diverse orbital structures, however. Dark matter haloes tend to be isotropic in the central regions and moderately tangential at large radii \citep{Taylor2001, Diemand2004, Ludlow2011, Svensmark2021}; for galaxies and globular clusters the orbital structure shows a larger spread in anisotropy \citep{vandeVen2006, Napolitano2014, Liepold2020, Santucci2022, Cohen2021}. For this reason it is useful to investigate to what extent our results can be generalised to more complex orbital structures. 

The isotropic orbital structure can be considered as a special case of a more general one-parameter family of orbital structures, called the Osipkov--Merritt orbital structures \citep{Osipkov1979, Merritt1985a}. These models are characterised by a distribution function that depends on binding energy and angular momentum only through the combination $Q = \calE - \tfrac12\,\lambda\,L^2$. The velocity anisotropy profile, generally defined as 
\begin{equation}
\beta(r) = 1- \frac{\sigma_\theta^2}{\sigma_r^2},
\end{equation}
has a specific form for the Osipkov--Merritt models:
\begin{equation}
\beta(r) = \frac{\lambda\,r^2}{1 + \lambda\,r^2}.
\end{equation}
For $\lambda = 0$, $Q$ reduces to $\calE$ and the Osipkov--Merritt orbital structure reduces to the isotropic one ($\beta = 0$). For $\lambda > 0$, the orbital structure varies from isotropic in the centre to purely radial ($\beta\to1$) at very large radii. This type of Osipkov--Merritt model is called Type~I in \citet{Merritt1985a} and is commonly adopted in the literature \citep[e.g.][]{Merritt1985b, Carollo1995, Ciotti1997, Widrow2000, Lokas2001, DiCintio2015, DiCintio2020, Ciotti2019}. Finally, for $\lambda < 0$, the orbital structure gradually changes from isotropic in the centre to completely tangential ($\beta\to-\infty$) at $r = r_{\text{a}} \equiv 1/\sqrt{-\lambda}$. These so-called Type~II Osipkov--Merritt models are only physically meaningful for spherical models with a finite extent with $r_{\text{a}} \leqslant \rT$, or equivalently with $\lambda \geqslant \lambda_{\text{min}} = -1/\rT^2$. Owing to this limitation, tangential Osipkov--Merritt models have only been considered in a limited number of studies, such as \citet{Polyachenko1974}, \citet{Osipkov1979}, and the previous papers in this series (\citetalias{Baes2022b}--\citetalias{Baes2024c}). 

The consistency conditions for any Osipkov--Merritt model depend on the value of $\lambda$. In general, it is easier to populate a given density distribution with nearly circular orbits than with very eccentric orbits. Therefore, if an Osipkov--Merritt model is consistent for a given value of $\lambda_0$, it will be consistent for all $\lambda$ in the range $\lambda_{\text{min}} \leqslant \lambda \leqslant \lambda_0$. To investigate the consistency of the family of softly truncated Hernquist models with an Osipkov--Merritt orbital structure, we hence have to find, for every $\rT$ and $\xi$, the maximum value of $\lambda$ for which the distribution function remains nonnegative for all $Q$. This is equivalent to finding, for every $(\rT,\lambda)$, the critical truncation sharpness $\xi_{\text{crit}}(\rT, \lambda)$.

Since the {\tt{SpheCow}} code can evaluate the Osipkov--Merritt distribution function for any spherical model \citep[for details, see][]{Baes2021b, Baes2023a}, we can numerically calculate $\xi_{\text{crit}}(\rT, \lambda)$ for each $(\rT,\lambda)$ combination. The result is shown in Fig.~{\ref{xicrit.fig}} for three different values of $\rT$. At fixed $\rT$, $\xi_{\text{crit}}$ is a decreasing function of $\lambda$, which implies that models with more tangential orbital structures allow sharper truncations. This is a logical consequence of the fact that it is easier to populate spherical models with nearly circular than with highly eccentric orbits. In the limit $\lambda\rT^2 = -1$ or $\lambda = \lambda_{\text{min}}$, we find $\xi_{\text{crit}} = 1$ for all $\rT$, which means that all truncations are allowed, including the infinitely sharp truncation. An interesting observation is that smoothly truncated models allow a range of orbital structures, including radially anisotropic Osipkov--Merritt models. This is again a stark difference from sharply truncated models, where only the extreme tangential Osipkov--Merritt model with $\lambda = -1/\rT^2$ is consistent \citepalias{Baes2023a}. 

As a concrete example, we consider the truncated Hernquist model with $\rT = 4$ and $\xi = 0.6$. For this model, all Osipkov--Merritt models with $-0.0625 \leqslant \lambda \leqslant 0.1828$ are consistent. On the bottom row of Fig.~{\ref{DFTP.fig}} we show the distribution function for the most tangential, the isotropic, and the most radial models in this range. The distribution functions are shown in turning-point space, which allows for an easier comparison than when shown as a function of $Q$, since the quantity $Q$ depends explicitly on the value of $\lambda$. The top row of this figure shows the distribution function of the corresponding non-truncated Hernquist models (only for the isotropic and radial models, as tangential Osipkov--Merritt models are not physically meaningful for models with an infinite extent).

Looking at the original Hernquist model, the most obvious difference between the isotropic and the radial Osipkov--Merritt model is the coverage of the allowed orbital space: in the isotropic case, all orbits are populated, whereas in the radial Osipkov--Merritt case the region in orbital space corresponding to $Q<0$ is unpopulated. Furthermore, the change in orbital structure changes the orientation of the iso-density contours: the effect is negligible in the central region, but at larger radii the orientation is slightly different, with flatter (steeper) contours for the isotropic (radial) model. Now looking at the bottom row, we see that the effect of the truncation is to remove all orbits with apocentre larger than the truncation radius. In the isotropic case (bottom-central panel), the truncation also creates a region of unpopulated, weakly bound orbits. In the case of the most tangential model (bottom-left panel) all orbits are populated, whereas in the most radial model (bottom-right panel) the populated region in turning-point space is even more reduced. In all cases, the bump--dip combination discussed in Sect.~{\ref{DynamicalProperties.sec}} is visible as the darker diagonal band in the distribution function. In the case of the most extreme radial Osipkov--Merritt model, the distribution function reaches zero at the location of the dip.

The key aspect of Fig.~{\ref{DFTP.fig}} is that, depending on the values of $\rT$ and $\xi$, softly truncated Hernquist models can accommodate different Osipkov--Merritt orbital structures, from extremely tangential over isotropic to even radial. This is a strong difference from sharply truncated models, which allow only the extreme tangential Osipkov--Merritt model \citepalias{Baes2023a}.

\subsection{Stability considerations}

An important caveat concerns the dynamical stability of the softly truncated models presented in this paper. Throughout this work we have focused on phase-space consistency, i.e.\ the requirement that the distribution function be non-negative everywhere. While this condition is necessary for a physical equilibrium, it is not sufficient to guarantee stability. For spherical systems with an isotropic distribution function, Antonov’s classical theorem states that a monotonic decrease of the distribution function with binding energy, $\txd f/ \txd\calE > 0$, is a sufficient but not necessary condition for stability against all perturbations \citep{Antonov1962, Doremus1971, Gillon1980, Perez1996, Binney2008}. The softly truncated models considered here generically violate this monotonicity condition through the characteristic bump--dip feature near the truncation energy, even when the distribution function remains everywhere positive.

Recent work has shown that such local inflections in the distribution function can have important dynamical consequences. \citet{Weinberg2023} demonstrated that spherical systems with truncated outer density profiles can develop an inflection in $f(\calE)$ that destabilises an otherwise weakly damped dipole ($\ell = 1$) mode, leading to a slowly growing lopsided instability. \citet{Dattathri2025} extended this analysis to a broad class of spherical double power-law models and showed, using both linear response theory and high-resolution $N$-body simulations, that isotropic systems with a pronounced bump in their distribution function generically develop a growing dipole mode. This instability saturates into a long-lived, off-centred configuration that sloshes through the central regions, while resonant interactions gradually erode the bump in phase space. 

The distribution functions of the softly truncated models constructed in this paper exhibit precisely the type of non-monotonic behaviour identified in these studies. Although the bump--dip feature arises here from an explicitly smooth radial truncation rather than from a sharp cutoff or incomplete mixing, its qualitative structure is similar. This suggests that, despite being phase-space consistent, some softly truncated isotropic models may be susceptible to dipole instabilities on secular timescales. Such an instability would not invalidate the models as instantaneous equilibria, but it would imply that they represent metastable configurations whose long-term evolution may involve slow reorganisation of the central structure.

The situation becomes richer for anisotropic distribution functions. Strong radial anisotropy is known to trigger the radial-orbit instability \citep{Antonov1973, Henon1973, Polyachenko1981, Polyachenko2015, Merritt1985c, Barnes1986}, and numerous studies have quantified stability limits for Osipkov--Merritt models in terms of the anisotropy radius or global anisotropy parameters \citep[e.g.][]{Palmer1987, Saha1991, Nipoti2002, DiCintio2020}. In addition, anisotropic spherical systems generically support weakly damped global modes \citep{Weinberg1991, Weinberg1994, Heggie2020}, which can be destabilised by relatively small features or inflections in the distribution function. Recent work has shown that truncation-induced bumps in Osipkov--Merritt models can excite slowly growing dipole modes through resonant couplings between inner and outer orbital families \citep{Weinberg2023, Dattathri2025}.

In this context, the consistency boundaries derived in Sect.~{\ref{OM.sec}} and illustrated in Fig.~{\ref{xicrit.fig}} should be interpreted with care. These boundaries delineate the region of parameter space in which the distribution function remains non-negative for a given truncation sharpness and anisotropy parameter. They do not, however, mark a transition between stable and unstable models. In particular, models that lie safely within the consistent region may still violate Antonov-type monotonicity conditions or host phase-space inflections capable of exciting secular instabilities. Soft truncations substantially enlarge the parameter space of consistent Osipkov--Merritt models compared to sharply truncated systems, but consistency alone does not preclude the development of dipole modes or, for sufficiently radial anisotropy, radial-orbit instabilities. We therefore caution the reader that the existence of a positive and well-behaved distribution function does not, by itself, guarantee long-term stability. 

A definitive assessment of the stability of softly truncated models, both isotropic and anisotropic, requires a dedicated investigation using linear stability analysis and/or controlled $N$-body simulations with initial conditions drawn directly from the distribution functions presented here. Such a study would allow one to quantify growth rates, identify the dominant modes, and explore how stability depends on truncation radius, truncation sharpness, and orbital anisotropy. This lies beyond the scope of the present paper, whose primary aim is to establish phase-space consistency and to provide a general framework for constructing finite-extent equilibrium models.

\subsection{The distribution function as consistency diagnostic}

An important aspect of the study presented here is the consistency of dynamical models. The only condition for a dynamical model to be consistent is that the distribution function is nonnegative over the entire phase space. In many cases, the calculation of the distribution function is far from trivial. In fact, the isotropic and Osipkov--Merritt orbital structures that we have discussed here are among the few cases where the calculation of the distribution function can be carried out using a relatively simple integration formula. In the general case, the inversion involves Laplace--Mellin integral transforms and is numerically unstable \citep{Dejonghe1986}. 

In light of this situation, it would be useful to have at one's disposal means of testing the consistency of dynamical models without the need to actually calculate the distribution function. Unfortunately, such general consistency criteria are not at hand. Necessary and sufficient conditions for phase-space consistency for special classes of spherical models were derived and applied by \citet{Ciotti1992}, \citet{Ciotti1996, Ciotti1999b}, \citet{Ciotti2010} and \citet{An2012}, but these conditions essentially only reduce the potential parameter space of consistent models. To derive the actual limits on phase-space consistency, a numerical evaluation of the distribution function remains unavoidable. 

The cases studied in this paper confirm and nicely illustrate this situation. In Table~{\ref{xicrit.tab}} and Fig.~{\ref{xicrit.fig}} we have derived consistency limits for a number of softly truncated Hernquist models, but this was only possible by numerically evaluating the distribution function. Looking back at Figs.~{\ref{Hernquist_basic.fig}} and~{\ref{Hernquist_iso.fig}}, we note that consistency or non-consistency of the ergodic dynamical models does not show up in any of the radial profiles shown. For example, both consistent and inconsistent dynamical models have dispersion profiles that look perfectly reasonable, and it is impossible to deduce the consistency of the models based on the basic properties and the dispersion profiles alone. This illustrates the need for methods to efficiently and accurately evaluate the distribution function of dynamical models. The {\tt{SpheCow}} code is essentially designed to do exactly this for models with an arbitrary density or surface density profile and has proven particularly useful. In summary, while various auxiliary criteria can provide useful guidance, point-wise inspection of the distribution function remains the only reliable test of phase-space consistency for dynamical models.

\section{Summary}
\label{summary.sec}

In this paper we have introduced a general method to construct self-consistent spherical dynamical models with a finite spatial extent by applying a smooth radial truncation to an arbitrary density profile. The truncation is implemented through a truncation function $\calS(x)$ that drives the density continuously to zero at the outer radius $\rT$, thereby avoiding the phase-space singularities that arise in sharply truncated models. This prescription has been incorporated in the {\tt{SpheCow}} code and applied to the Hernquist model as a representative example.

For isotropic models, we find that softly truncated Hernquist models are dynamically consistent for a wide range of truncation radii, provided that the truncation is sufficiently gradual. The corresponding distribution functions exhibit a characteristic bump--dip structure near the truncation energy $\calET$, and the depth of this dip yields a simple diagnostic for identifying the critical truncation sharpness $\xi_{\text{crit}}(\rT)$.

We have also examined truncated models with an Osipkov--Merritt orbital structure. In contrast to sharply truncated systems, which only admit the extreme tangential limit, softly truncated models accommodate a broad family of anisotropic equilibria, including moderately radial ones. For each $(\rT,\lambda)$ we determined the maximum truncation sharpness $\xi_{\text{crit}}(\rT,\lambda)$ that preserves phase-space consistency. The resulting consistency landscape demonstrates that soft truncations dramatically enlarge the parameter space of dynamically viable finite-extent models.

Soft radial truncations therefore provide a physically motivated and versatile framework for constructing finite-extent dynamical models with controlled outer boundaries. The approach is general and can be applied to any spherical density profile, offering a useful tool for analytical studies, dynamical modelling, and the generation of equilibrium $N$-body initial conditions. Although the softly truncated models presented here are phase-space consistent, their long-term dynamical stability is not guaranteed: the characteristic bump--dip feature in the distribution function may render some models susceptible to slow, secular instabilities, an issue that warrants further study using linear stability analyses or $N$-body simulations.

\section*{Data availability}

No astronomical data were used in this research. The data generated and the plotting routines will be shared on reasonable request to the corresponding author. The {\tt{SpheCow}} code is publicly available on GitHub.

\section*{Acknowledgements}

The author is grateful to the anonymous referee for a thoughtful and constructive report that led to several improvements in the manuscript.

\bibliographystyle{mnras}
\bibliography{../../../MyBib/mybib_nameyear}

\appendix

\section{Alternative choices for the truncation function}
\label{appendix.sec}

In Sect.~{\ref{Math.sec}} we introduced a smooth truncation function $\calS(x)$ based on the cumulative distribution function of a logit--normal distribution. This choice ensures that the density and all of its derivatives vanish at the truncation radius $\rT$, thereby preventing the phase-space singularities that arise for sharply truncated models. The specific form of $\calS(x)$ is not unique, however: any sufficiently smooth function that transitions monotonically from unity to zero may be used. Ideally, $\calS(x)$ is infinitely differentiable ($C^\infty$); it should at minimum be twice differentiable ($C^2$) for the Eddington inversion to remain well behaved. 

One alternative with specific control over smoothness is the use of compactly supported polynomials. Polynomial transition functions of the so-called smoothstep family are commonly used in computer graphics and machine learning \citep{Perlin1985, Ebert2003}. The general form for the $n$th smoothstep function is given by 
\begin{equation}
\calS_n(x) = \begin{cases}
\; 0 & \quad{\text{for }}x\leqslant 0, \\
\displaystyle
\; x^{n+1} \sum_{k=0}^n \begin{pmatrix} n+k \\ k \end{pmatrix} 
\begin{pmatrix} 2n+1 \\ n-k\end{pmatrix} (-x)^k
& \quad{\text{for }}0\leqslant x \leqslant 1, \\
\; 1 & \quad{\text{for }}x\geqslant 1,
\end{cases}
\end{equation}
Higher values of $n$ correspond to smoother polynomials at the end points. Once $n\geqslant2$, the first and second derivatives are zero at the edges of the interval. For our purposes, one would typically adopt $n\geqslant2$ so that $\rho(r)$ is at least $C^2$ at the truncation radius.

Wendland functions form a family of compactly supported radial basis functions with minimal polynomial degree for a prescribed level of smoothness \citep{Wendland1995}. They have become standard tools in scattered-data approximation and meshless numerical methods \citep[e.g.][]{Schaback2007, Fasshauer2012}, and are widely used as smoothing kernels in smoothed particle hydrodynamics \citep{Dehnen2012, Wissing2025}. In \citetalias{Baes2024c} we already employed Wendland functions as the basis for dynamical models with a finite extent, and their mathematical properties make them equally suitable as truncation functions. In three dimensions, each Wendland function, denoted as $\phi_{k+2,k}(u)$, is a monotonically decreasing $C^{2k}$ polynomial in $u$ of order $3k+2$ with $\phi_{k+2,k}(0) = 1$ and $\phi_{k+2,k}(1) = 0$. For any value of $k$, the function  
\begin{equation}
\calS_k(x) = \begin{cases}
\; 0 & \quad{\text{for }}x\leqslant 0, \\
\; 1 - \phi_{k+2,k}(x) & \quad{\text{for }}0\leqslant x \leqslant 1, \\
\; 1 & \quad{\text{for }}x\geqslant 1,
\end{cases}
\end{equation}
satisfies the required conditions for the truncation function. Wendland-based tapers thus offer an attractive alternative to the logit--normal truncation function: they provide exact compact support, explicit polynomial structure, and full control over the degree of differentiability.

\bsp	
\label{lastpage}
\end{document}